\documentclass[11pt,a4paper]{article}

\usepackage[colorlinks=true, linkcolor=black!50!blue, urlcolor=blue, citecolor=blue, anchorcolor=blue]{hyperref}
\usepackage[font=small,labelfont=bf,margin=0mm,labelsep=period,tableposition=top]{caption}
\usepackage[a4paper,top=3cm,bottom=2.5cm,left=2.5cm,right=2.5cm,bindingoffset=0mm]{geometry}

\usepackage{graphicx}
\usepackage{float}
\usepackage{afterpage}
\usepackage{epsfig,cite}
\usepackage{amssymb}
\usepackage{amsmath}
\usepackage{bm}
\usepackage{dsfont}
\usepackage{multirow}
\usepackage{url}
\usepackage{xcolor}
\usepackage{float}
\usepackage{afterpage}
\usepackage{ulem}
\usepackage{gensymb}
\usepackage{url,inputenc}
\usepackage{hyperref}
\usepackage{placeins}
\usepackage{indentfirst}
\usepackage{multirow,booktabs,multirow}

\usepackage{amsmath} 
\newcommand{\angstrom}{\textup{\AA}}

\newcommand{\be}{\begin{equation}}
\newcommand{\ee}{\end{equation}}
\newcommand{\bea}{\begin{eqnarray}}
\newcommand{\eea}{\end{eqnarray}}
\newcommand{\bi}{\begin{itemize}}
\newcommand{\ei}{\end{itemize}}
\newcommand{\ben}{\begin{enumerate}}
\newcommand{\een}{\end{enumerate}}

\def\frac#1#2{{{#1}\over {#2}}}
\def\gsim{\mathrel{\rlap{\lower4pt\hbox{\hskip1pt$\sim$}}
    \raise1pt\hbox{$>$}}}       
\def\lsim{\mathrel{\rlap{\lower4pt\hbox{\hskip1pt$\sim$}}
    \raise1pt\hbox{$<$}}}

\newcommand{\draft}[1]{}

\def\beq{\begin{equation}}
\def\eeq{\end{equation}}

\def\lapprox{\lower .7ex\hbox{$\;\stackrel{\textstyle <}{\sim}\;$}}
\def\gapprox{\lower .7ex\hbox{$\;\stackrel{\textstyle >}{\sim}\;$}}

\usepackage{tabularx}
\newcolumntype{C}[1]{>{\centering\arraybackslash}p{#1}}

\begin{document}
\newgeometry{top=1.5cm,bottom=2.5cm,left=2.5cm,right=2.5cm,bindingoffset=0mm}

$\quad$
\vspace{1.5cm}

\begin{center}

{\LARGE \bf Metallic Edge States in Zig-Zag  \\ \vspace{0.2cm} Vertically-Oriented MoS$_2$ Nanowalls} \\

      \vspace{1.4cm}

M. Tinoco$^{1,2,\dagger}$, L. Maduro$^{1,\dagger}$ and S. Conesa-Boj$^{1,*}$ \vspace{0.6cm}
       
	{\it ~$^{1}$ Kavli Institute of Nanoscience, Delft University of Technology,\\ 2628CJ Delft, the Netherlands.\vspace{0.2cm}

	~$^{2}$ ICTS – Centro Nacional de Microscop\'ia Electr\'onica,\\ Universidad Complutense, 28040 Madrid, Spain.}\\
      
      \vspace{1.4cm}
      
  
      {\bf \large Abstract}
      
      \end{center}
      
The remarkable properties of layered materials such as MoS$_2$ strongly depend on their dimensionality. Beyond manipulating their dimensions, it has been predicted that the electronic properties of MoS$_2$ can also be tailored by carefully selecting the type of edge sites exposed. However, achieving full control over the type of exposed edge sites while simultaneously modifying the dimensionality of the nanostructures is highly challenging. Here we adopt a top-down approach based on focus ion beam in order to selectively pattern the exposed edge sites. This strategy allows us to select either the armchair (AC) or the zig-zag (ZZ) edges in the MoS$_2$ nanostructures, as confirmed by high-resolution transmission electron microscopy measurements. The edge-type dependence of the local electronic properties in these MoS$_2$ nanostructures is studied by means of electron energy-loss spectroscopy measurements.
This way, we demonstrate that the ZZ-MoS$_2$ nanostructures exhibit clear fingerprints of their predicted metallic character. Our results pave the way towards novel approaches for the design and fabrication of more complex nanostructures based on MoS$_2$ and related layered materials for applications in fields such as electronics, optoelectronics, photovoltaics, and photocatalysts. \\

 \vspace{1.4cm}

 ~$^{*}$Corresponding author: \href{mailto:s.conesaboj@tudelft.nl }{s.conesaboj@tudelft.nl } \vspace{0.1cm}

~$^{\dagger}$ Equal contribution.

\clearpage

\section*{Introduction}

The ability of crafting new materials in a way that makes possible controlling and enhancing their properties is one of the main requirements of the ongoing nanotechnology revolution~\cite{ref1,ref2}. In this context, a family of materials that has attracted intense attention recently are 2D layered materials, such as MoS$_2$, which belong to the group of transition metal dichalcogenides (TMDs). These materials have been extensively studied due to their promising electrical and optical properties~\cite{ref3,ref4,ref5,ref6}. A defining feature of TMDs is that they exhibit a lack of inversion symmetry, which leads to the appearance of a variety of different edge structures. The most common of these, consisting on dangling bounds, are the armchair (AC) and the zig-zag (ZZ) edge structures.

\vspace{0.2cm}

Of particular relevance in this context, the electronic properties of MoS$_2$ have been predicted to be affected by the presence of the different edge structures in rather different ways. For instance, the AC edges have been predicted to be semiconducting, while the ZZ edges should exhibit instead metallic behavior~\cite{ref7,ref8,ref9,ref10}. Moreover, {\it ab-initio} theoretical calculations predict that these metallic states at the edges of MoS$_2$ could lead to the formation of plasmons~\cite{ref11}. Beyond this tuning of electronic properties, other attractive applications of these active edge sites arise in photocatalysis, such as their use in hydrogen evolution reactions (HER)~\cite{ref12,ref13,ref14,ref15}.

\vspace{0.2cm}

With these motivations, it is clear that the design and fabrication of MoS$_2$ nanostructures with morphologies that maximize the number of exposed active edge sites is a key aspect for further improvements in terms of applications. In this respect, significant efforts have been pursued to realize the systematic bottom-up growth of vertically-oriented standing MoS$_2$ layers. This configuration leads to the edge sites facing upwards, therefore maximizing the number of exposed edge sites as compared with the more common horizontal configuration, where its basal plane lies parallel to the substrate~\cite{ref16,ref17,ref18,ref19,ref20,ref21,ref22,ref23}. However, this bottom-up approach is hampered by a lack of reproducibility due to the complexities of the growth mechanism. Another limitation within the bottom-up approach is that the specific type of edges exposed cannot be selectively grown.

\vspace{0.2cm}

Ideally, one would like to combine the best of both worlds. On the one hand, it is important to be able to controllably grow MoS$_2$ nanostructures that exhibit the largest possible surface area of edge structures, as it is achieved by the bottom-up strategy summarized above. On the other hand, one would also like to be able to select the specific type of edge sites exposed, in particular, by selecting whether these correspond to AC or to ZZ edges. Therefore, the main goal of this work is to bridge these two requirements by realizing a novel approach to the growth of vertically-oriented standing MoS$_2$ layers with full control on the nature of the exposed edge sites.

\vspace{0.2cm}

To achieve this goal, here we adopt a well-stablished top-down approach based on focus ion beam (FIB) in a way that allows us to selectively pattern both types of edges (AC and ZZ) within out-of-plane (vertical) MoS$_2$ nanostructures. In the context of patterning layered materials, the usefulness of FIB has been repeatedly demonstrated.~\cite{ref24,ref25,ref26}. By means of this technique, we are able to selectively maximize the density of exposed edge sites while controlling their type. Subsequently, by combining high-resolution transmission electron microscopy (TEM) with electron energy-loss spectroscopy (EELS) measurements, we are able to confirm not only the crystallographic nature of both the AC and ZZ MoS$_2$ surfaces, but also we can demonstrate that, despite the roughness and imperfections induced during the fabrication procedure, the ZZ MoS$_2$ nanostructures clearly exhibit a metallic character, in agreement with the theoretical predictions from {\it ab-initio} calculations~\cite{ref11}.

 \vspace{0.2cm}

The results of this work will open new opportunities for nanoengineering the edge type in MoS$_2$ nanostructures as well as in related layered materials, paving the way towards novel exciting opportunities both for fundamental physics and technological applications in electronics, optoelectronics, photovoltaics, and photocatalysts.

\section*{Results}

\begin{figure}[h]
  \begin{center}
  \includegraphics[width=0.80\textwidth]{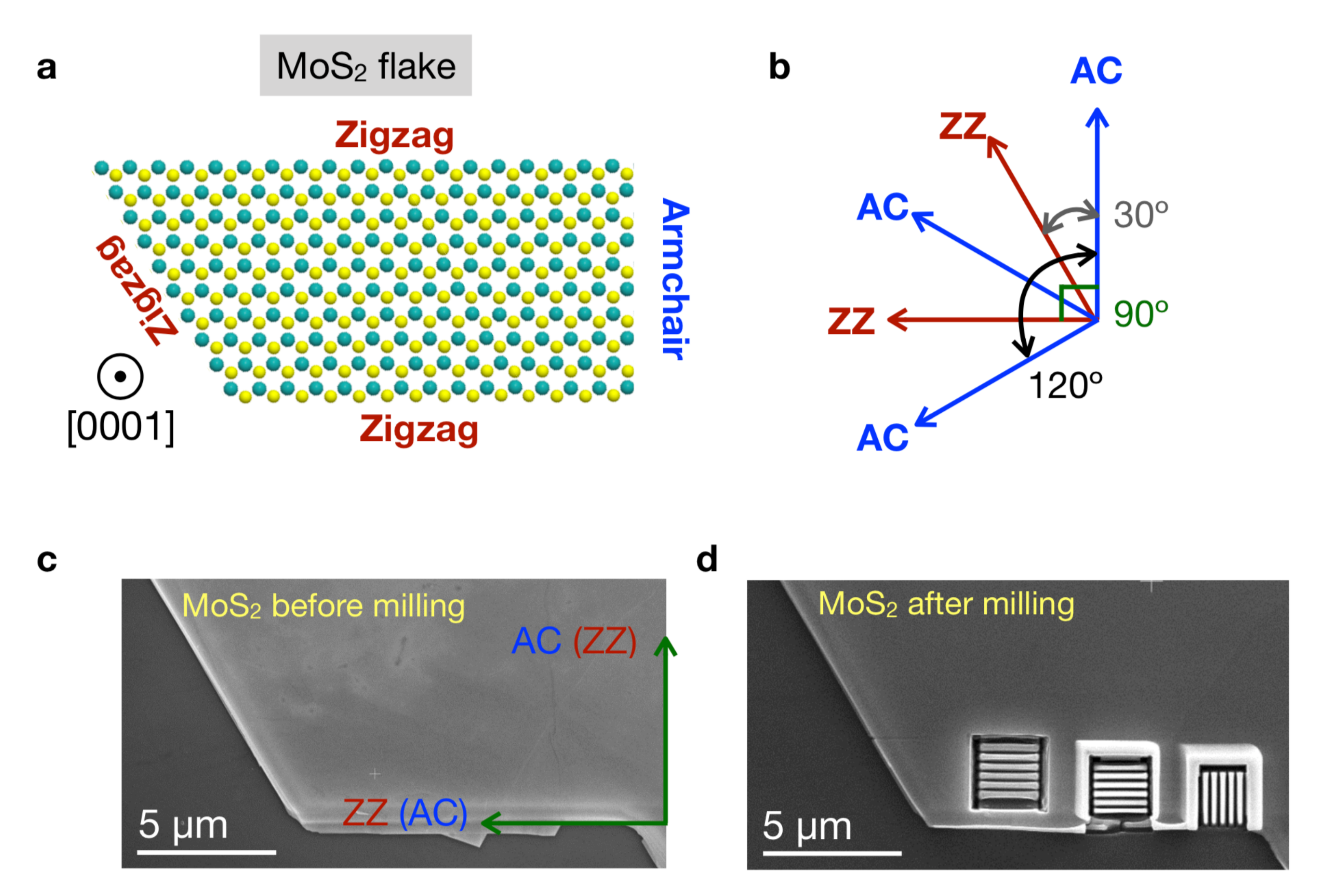}
  \caption{\small (a) Atomic model of a MoS$_2$ flake viewed along the [0001] direction, where we indicate the corresponding zig-zag (ZZ) and armchair (AC) edges. (b) From geometric considerations, we can determine the possible values that the angles between adjacent AC and ZZ edges should take; (c) and (d) SEM micrographs of the MoS$_2$ flake used for patterning the nanowalls, taken before and after the milling respectively. In (d), three different set of arrays can be observed. The left-most array was fabricated without the protective metal layer, while the other arrays used instead this protective metal layer.
    \label{fig1} }
  \end{center}
\end{figure}

From crystal structure considerations, the possible angles between adjacent flat edges within MoS$_2$ flakes should be multiples of 30\degree. Specifically, the expected angles between adjacent AC and ZZ edge structures in a MoS$_2$ flake such as that of {\bf Fig.~\ref{fig1}a}  should be 30\degree, 90\degree, and 120\degree, as illustrated in {\bf Fig.~\ref{fig1}b}. Based on this information, we have designed the orientation of the different areas of the MoS$_2$ flake that subsequently will be patterned. In this way, we can ensure the full control over the resulting specific edge crystallographic orientation.

 \vspace{0.2cm}

{\bf Figs.~\ref{fig1}c} and{\bf~\ref{fig1}d} display a scanning electron microscopy (SEM) image of the MoS$_2$ flake that has been used for the fabrication of the nanostructures, taken before and after the milling respectively. Before the milling is performed, a protective metallic layer of tungsten (W) with a thickness of 500 nm was deposited on top of the selected areas of the MoS$_2$ flake. Subsequently, we performed a series of milling and cleaning processes in order to construct the vertically-aligned MoS$_2$ nanostructures. {\bf Fig.~\ref{fig1}d} displays three ordered vertically-oriented patterned arrays of MoS$_2$ nanostructures, which in the following are denoted as nanowalls (NWs). Two of these sets of nanowalls are oriented perpendicularly with respect to each other, guaranteeing that this way one of two arrays will correspond to AC (ZZ) NWs while the other array will correspond instead to the complementary ZZ (AC) ones. These NWs are found to exhibit a uniform thickness being ($89 \pm  5$) nm (central array in {\bf Fig.~\ref{fig1}d}) and ($68 \pm 5$) nm (rightmost array in {\bf Fig.~\ref{fig1}d}). Note that the left-most array was fabricated without the protective metal layer. 


 \vspace{0.2cm}

To further examine the crystallographic nature of the resulting vertical MoS$_2$ nanostructures, transmission electron microscopy (TEM) studies were carried out. For these studies, we lifted out two of the MoS$_2$ NWs from the two different patterned NWs arrays using a micromanipulator. Subsequently, the nanostructure was mounted onto a TEM half-grid. This whole procedure takes place within the FIB chamber.

\FloatBarrier

\clearpage

\begin{figure}[t]
  \begin{center}
  \includegraphics[width=0.48\textwidth,angle=-90]{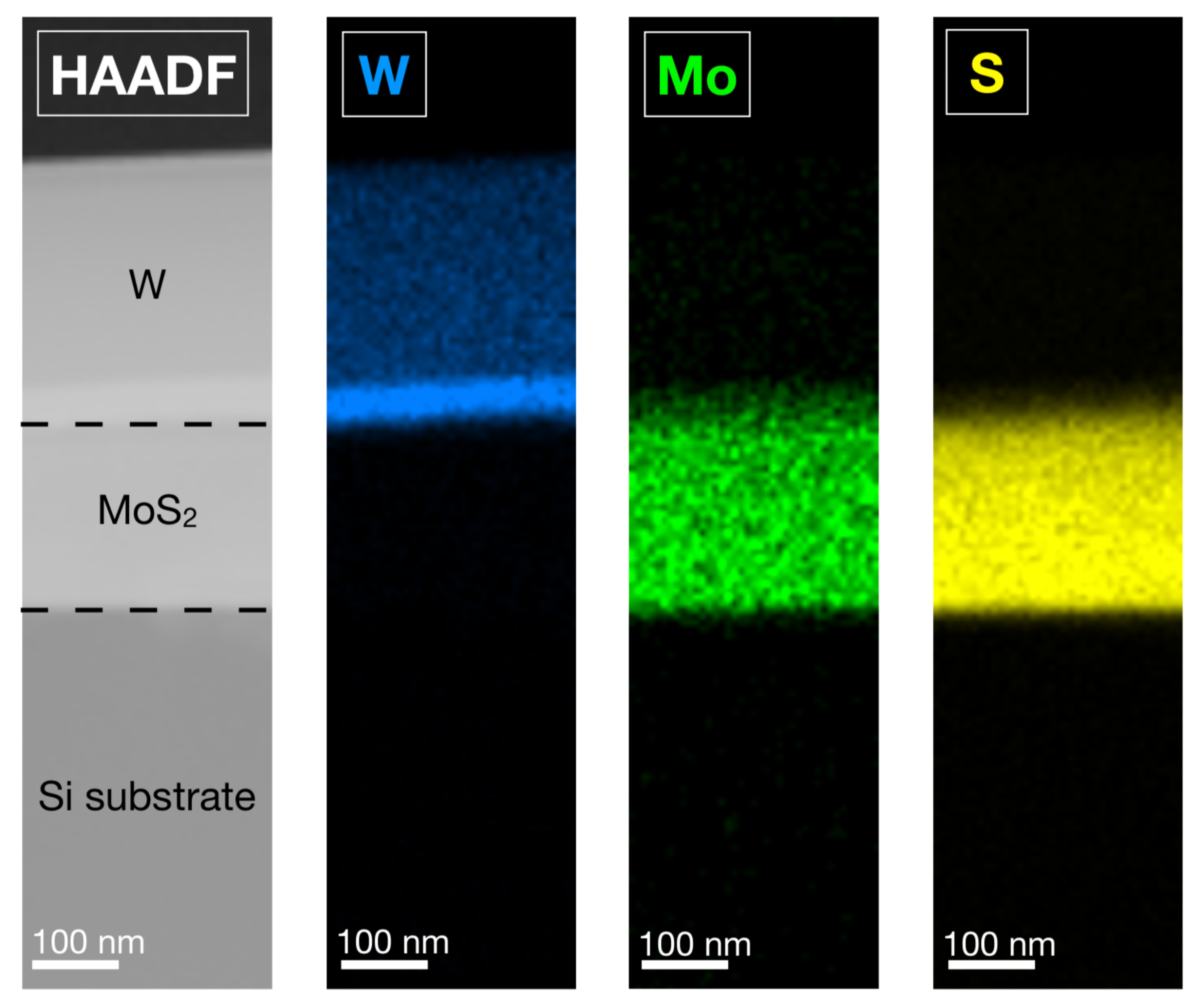}
  \caption{\small 
     (a) HAADF-STEM image of an area of the ZZ MoS$_2$ NW, which is bracketed between the Si substrate and the metallic protective layer; (b) The corresponding EDX compositional maps (Mo in green, S in yellow, W in blue).
     \label{fig2} }
  \end{center}
\end{figure}

{\bf Fig.~\ref{fig2}a} displays a high-angle annular dark-field scanning transmission electron microscopy (HAADF-STEM) image of a selected region of the ZZ MoS$_2$ nanowall, bracketed between the Si substrate and the metallic protective layer.
{\bf Fig.~\ref{fig2}b} shows the corresponding chemical compositional of this nanowall obtained by means of energy dispersive X-ray (EDS) spectroscopy measurements. From the EDS map, the different chemical components of the NWs can be clearly distinguished: the MoS$_2$ segment, embedded within the protective metal layer tungsten (W), and the silicon (Si) substrate.

 \vspace{0.2cm}

\begin{figure}[h!]
  \begin{center}
  \includegraphics[width=0.58\textwidth,angle=-90]{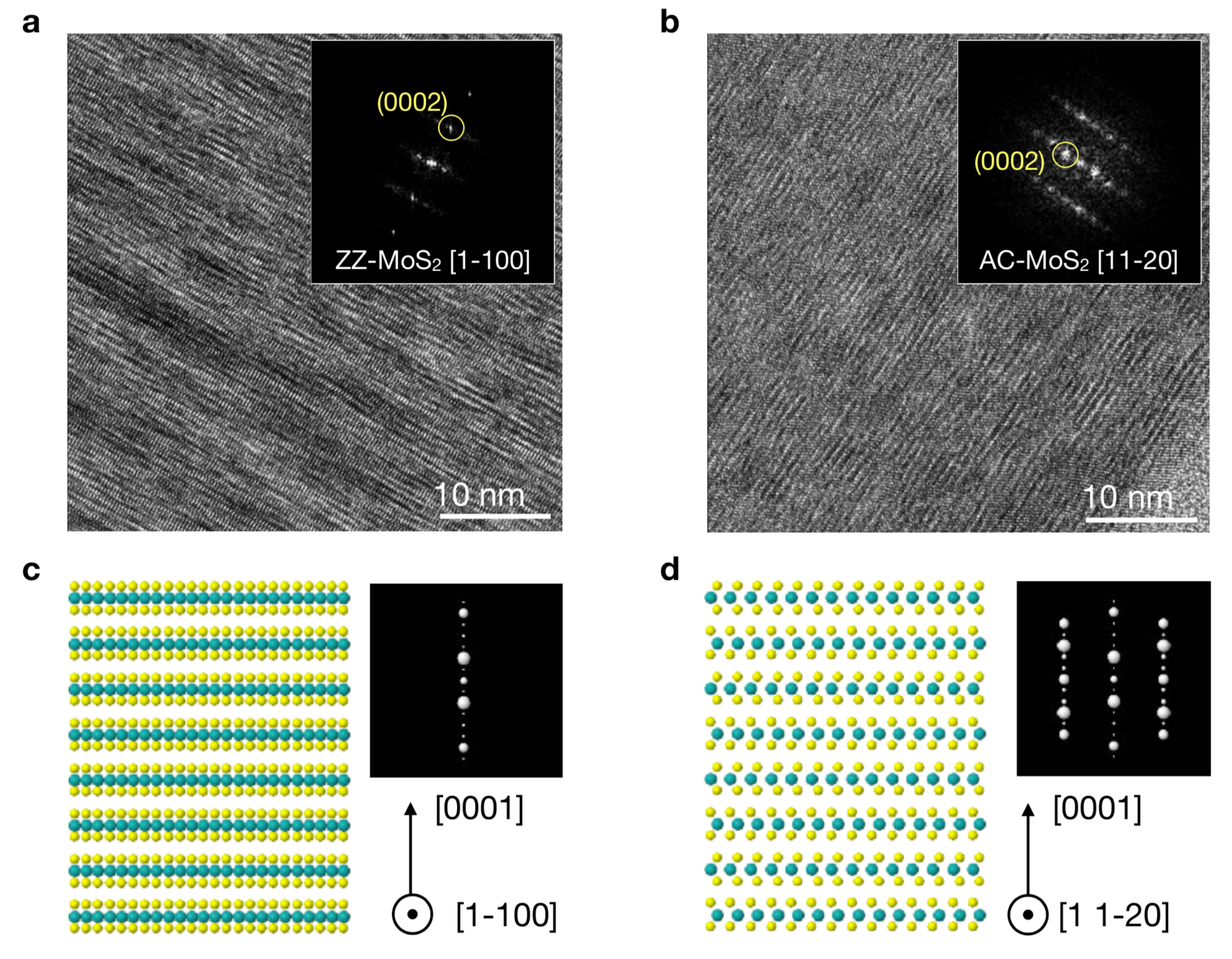}
  \caption{\small 
   (a) and (b) HRTEM micrographs of representative ZZ and AC MoS$_2$ nanowalls, respectively. The insets indicate the corresponding fast Fourier transform (FFT). (c) and (b) The atomic modelling associated to the AC and ZZ orientations of the NWs, together with the theoretical calculation of the expected FFTs.
     \label{fig3} }
  \end{center}
\end{figure}

From the crystalline structure studies carried out by means of high-resolution TEM measurements ({\bf Fig.~\ref{fig3}}), we are able to confirm the specific edge site configuration for the two NW arrays. {\bf Figs.~\ref{fig3}a} and{\bf~\ref{fig3}b} display the results of the TEM measurements on the AC and ZZ MoS$_2$ surfaces respectively. By comparing the two crystallographic orientations, AC and ZZ, we can observe the differences between the atomic arrangement of each surface, which are consequently characterized by different fast Fourier transforms (FFTs) (shown in the insets of {\bf Figs.~\ref{fig3}a} and{\bf~\ref{fig3}b}). From these results, it is clearly noticeable the excellent agreement between the experimental FFTs obtained from the TEM measurements and the corresponding ones calculated theoretically in terms of the expected atomic configuration (shown in {\bf Figs.~\ref{fig3}c} and{\bf~\ref{fig3}d}). These results provide direct confirmation that these vertically-oriented MoS$_2$ nanowalls are in fact exposing ZZ and AC edge terminations, therefore validating our fabrication strategy.

\subsection*{ Fingerprinting the edge-type nature of MoS$_2$ nanowalls}

\begin{figure}[h!]
  \begin{center}
  \includegraphics[width=0.40\textwidth,angle=-90]{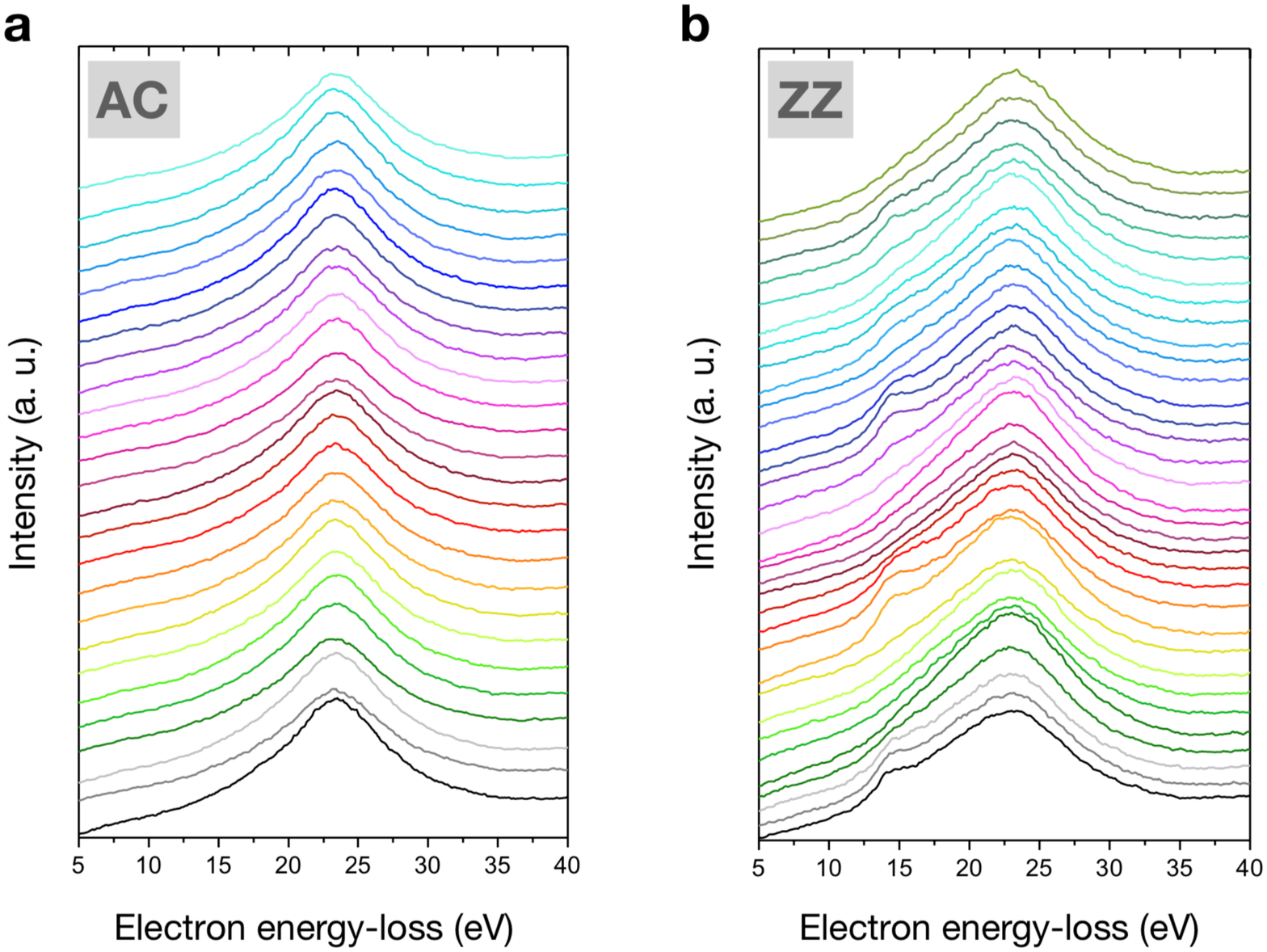}
  \caption{\small 
    (a) and (b) EELS spectra taken at different positions of the AC and ZZ nanowalls, respectively, for the region of electron energy losses between 5 and 40 eV. Each curve corresponds to a different position along the NWs.
      \label{fig4} }
  \end{center}
\end{figure}

In order to pin down the local electronic properties of the AC and ZZ MoS$_2$ NWs, electron energy-loss spectroscopy (EELS) measurements have been carried out in a scanning transmission electron microscope (STEM). In {\bf Fig.~\ref{fig4}} we show the energy-loss spectra corresponding to both the AC and ZZ surfaces, taken at different points along the length of the nanowall. As it can be observed in the two sets of EELS spectra, the MoS$_2$ bulk plasmon signal appears at 23.4 eV in both samples with similar intensities and general shape, in agreement with previous analyses ~\cite{ref27,ref28}. Nevertheless, the MoS$_2$ surface plasmon peak, present at 15.2 eV, turns out to appear only on a restricted subset of the spectra of the ZZ-nanowalls. Considering that the fabricated AC-nanowalls are thinner than the ZZ-terminated ones, the presence of the surface MoS$_2$ plasmon on the ZZ-nanowalls cannot be attributed to a lower thickness of the sample. Therefore, the origin of this peak should be caused by another phenomenon. In that respect, it is important to notice that the MoS$_2$ surface plasmon peak appears and disappears in a periodic manner, depending on the specific position along the nanowall where the EELS spectrum is collected. It is found that the positions which correspond to local maxima of the intensity associated to this surface plasmon peak are separated by around 12 nm between each other. This behavior can be attributed to the presence of metallic surface plasmon polaritons (SPP), which are planar waves appearing at the interfaces between a metal and a dielectric material under some external excitation, such as an electron beam~\cite{ref29}. That could correspond to the oscillatory character present in our EELS spectra. 

 \vspace{0.2cm}
 
Therefore, from this analysis, we can conclude that the ZZ MoS$_2$ NWs surfaces present a clear metallic character. On the contrary, the AC MoS$_2$ NWs do not exhibit such metallic behavior. With this result, we can hereby confirm that the ZZ MoS$_2$ NWs are dominantly enclosed by zig-zag edges structures. It is also worth mentioning here that no signal arising from neither the metal layer nor the Ga used for the FIB milling were present at any of the acquired EELS spectra, indicating that the possible contamination from Ga in the nanowalls is non-existing.

\section*{Discussion}

In order to further validate the onset of the metallic behavior observed in the ZZ MoS$_2$ nanowalls (NWs), we calculated the corresponding density of states (DOS) by means of {\it ab-initio} calculations in the framework of density functional theory (DFT). The van der Waals (vdW) interactions characteristic of MoS$_2$ were incorporated by using the nonlocal vdW functional model~\cite{ref30} as implemented in the WIEN2k code (see Methods for further details).

 \vspace{0.2cm}
 
\begin{figure}[h]
  \begin{center}
  \includegraphics[width=0.45\textwidth,angle=-90]{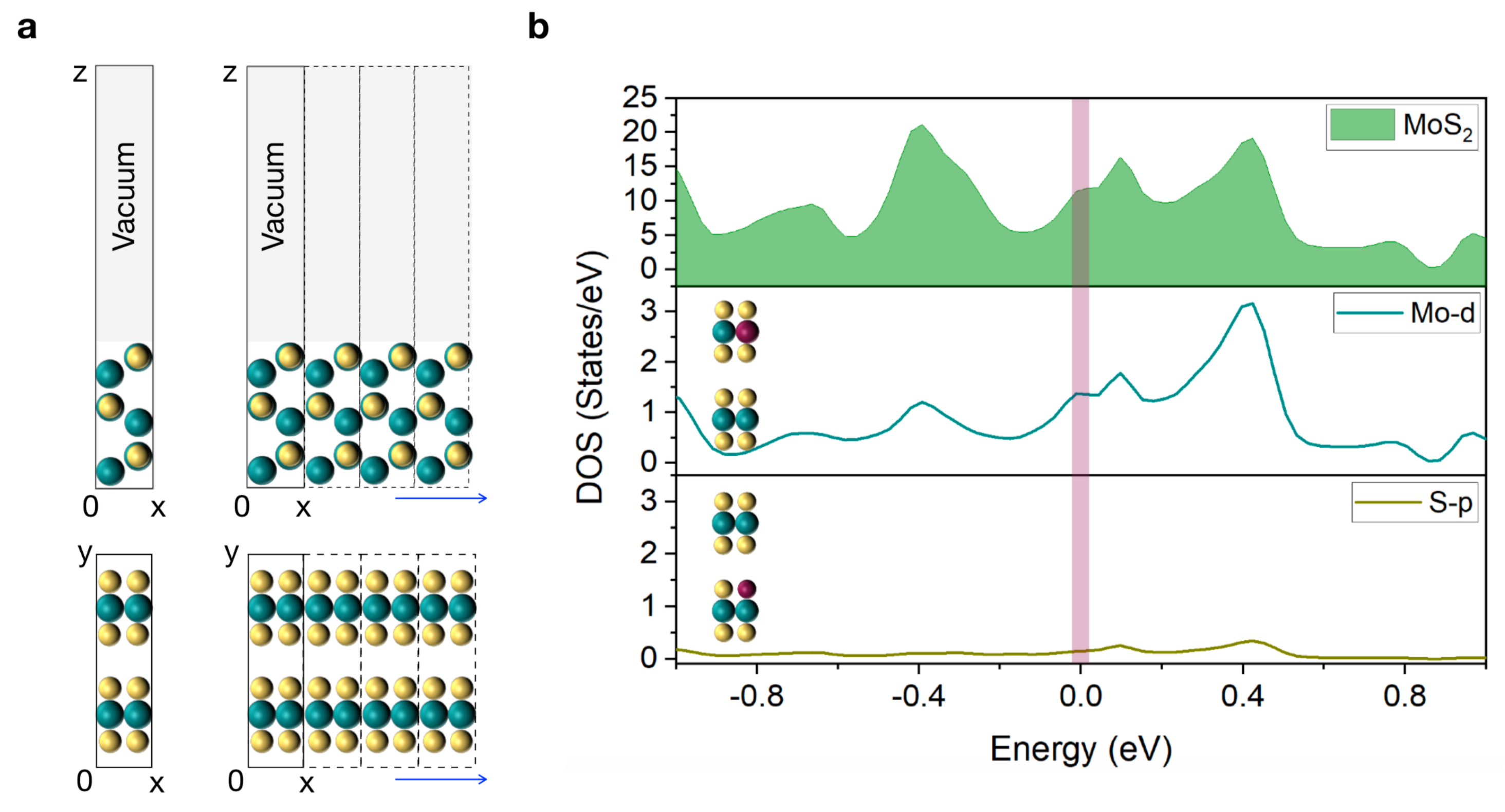}
  \caption{\small 
    (a) The ZZ MoS$_2$ nanowall can be modelled by constructing a 1x3x1 supercell with vacuum in the z direction. A vacuum layer with length of 17.170 $\angstrom$ was inserted along the ZZ edge of the nanowall, in order to to avoid spurious interactions between the repeating supercell images. This 1x3x1 supercell in the x and z directions was used to determine the density of states associated to the ZZ MoS$_2$ nanowall, which can be treated as a two-dimensional sheet composed by ZZ nanorribons stacked along the y direction. (b) (Top panel) density of states of ZZ MoS$_2$ nanowalls, (central and bottom panels) the individual contributions to the DOS from the Mo and S atoms located at the ZZ surface.
     \label{fig5} }
  \end{center}
\end{figure}

We modeled the ZZ MoS$_2$ nanowall by constructing a 1x3x1 supercell of MoS$_2$, as shown in {\bf Fig.~\ref{fig5}a}. In order to minimize the interactions between periodic images due to the 3D boundary conditions, we introduced a vacuum layer such that the distance between periodic images is 17.170 \angstrom. The resulting calculated total DOS for the ZZ MoS$_2$ NW is displayed in {\bf Fig.~\ref{fig5}b}.
A clear absence of a gap in the DOS near the Fermi energy is observed, which implies a finite probability (11.65 states/eV) for states just below and above the Fermi energy level being populated, highlighting the metallic behavior of the ZZ MoS$_2$ NWs. {\bf Fig.~\ref{fig5}b} (middle panel) also displays the individual contributions of the 4d states of Mo atoms located at the surface of the ZZ MoS$_2$ nanowall. These 4d states of Mo are also observed to cross over the Fermi energy (1.35 states/eV), contributing therefore to the metallic character of the ZZ MoS$_2$ nanowall. The individual contribution of the 3p states of S atoms located at the surface of the NW turns out to be much smaller from the DFT calculation, 0.08 states/eV. Therefore, the dominant contribution to the metallic character of ZZ MoS$_2$ nanowalls can be confidently attributed to Mo-4d states of MoS$_2$.

 \FloatBarrier
  \vspace{0.2cm}
 
In this work, we have presented a novel approach for the top-down fabrication of ordered vertically-oriented MoS$_2$ nanostructures (denoted as nanowalls) which makes possible to achieve at the same time, a large density of exposed active edge sites while also to controllably select whether these are of the AC or ZZ types. The crystallographic nature of the exposed surfaces has been validated by means of high-resolution TEM measurements. We have also studied the local electronic properties of these NW surfaces by means of EELS, finding direct evidence of the metallic character of the ZZ surfaces as indicated by the presence of MoS$_2$ surface plasmon peak.
 
 \vspace{0.2cm}

The metallic nature of the ZZ MoS$_2$ nanowalls can be exploited to open new opportunities for nanoengineering the edge type in MoS$_2$ nanostructures as well as in related layered materials. This would allow new exciting opportunities both for fundamental physics and technological applications in electronics, optoelectronics, photovoltaics, and photocatalysts.

\subsection*{Methods} 

	{\it Focus ion beam patterning for the fabrication of edge-controlled MoS$_2$ nanowalls}.
MoS$_2$ bulk crystal obtained from Alfa Aesar (99.999\% purity) was mechanical exfoliated with Poly-Di-Methyl-Siloxaan (PDMS) and then transferred to a SiO$_2$/Si substrate.
The MoS$_2$ nanostructures were milled using a FEI Helios G4 CX focus ion beam. The ion milling procedure was carried out using a very low energy electron beam of 15 kV, and an ion beam of 2 pA. Before the milling procedure was carried out, a protective metal (W) layer of 500 nm of thickness was deposited on top the selected areas.\\

{\it Characterization techniques}.
Transmission Electron Microscopy (TEM) measurements were carried out in a Titan Cube microscope using an acceleration voltage of 300 kV. Its spatial resolution at Scherzer defocus conditions is 0.08 nm in the High-Resolution Transmission Electron Microscopy (HRTEM) mode, whilst the resolution is around 0.19 nm in the HAADF-STEM (High Angle Annular Dark Field – Scanning Transmission Electron Microscopy) mode. Electron Energy Loss Spectroscopy (EELS) experiments were carried out using a Gatan Imaging Filter (GIF) spectrometer, employing a collection semi-angle of 2.95 mrad, a convergence semi-angle of 14 mrad, and an aperture of 2 mm.
The energy resolution obtained by using these parameters in EELS was 0.9 eV, with an exposure time of 0.1 s/spectrum and an energy dispersion of 0.1 eV/channel.\\

{\it First-principle calculations}.
The density of states (DOS) calculations were performed using both linearized augmented plane wave (LAPW) and local orbitals (LO) methods implemented in the WIEN2k package~\cite{ref31}.
The nonlocal van der Waals~\cite{ref32,ref33} (vdW) interactions used for the DOS calculations uses optB88~\cite{ref34} for the exchange term, the local density approximation~\cite{ref35} (LDA) for the correlation term, and the DRSLL kernel for the non-local term~\cite{ref36}.
For the non-local vdW integration the cut-off density rc was set to 0.3 bohr$^{-3}$, while the plane wave expansion cut-off Gmax was set to 20 bohr$^{-1}$. No spin polarization was considered. The lattice parameters were found by volume and force optimization of the supercell, such that the force on each atom was less than 1.0 mRy/bohr. The total energy convergence criteria was set to be 0.1 mRy between self-consistent field (SCF) cycles, while the charge convergence criteria was set to 0.001e, with e the elementary unit charge. The core and valence electron states were seperated by an energy gap of -6.0 Ry. Furthermore, the calculations used an R*kmax of 6.0, where R is the radius of the smallest Muffin Tin sphere, and kmax is the largest k-vector.
The first Brillouin zone for the lattice parameter calculations was sampled with 100 k-points using the tetrahedon method of Blöchl et  al.~\cite{ref37}, which corresponds to 21 k-points in the irreducible Brillouin zone. With the above parameters the optimized lattice parameters were $a = 3.107 \angstrom$
and $c = 12.087 \angstrom$, which are in good agreement with the experimental values
$a = 3.161 \angstrom$ and $c = 12.295 \angstrom$ ~\cite{ref38}.
The DOS was calculated with a denser k-point sampling of the Brillouin zone consisting of 1600 k-points, corresponding to 630 k-points in the irreducible Brillouin zone.

\subsection*{Acknowledgements}

M.~T and L.~M. acknowledge support from the Netherlands Organizational for Scientific Research (NWO) through the NanoFront program.
S.~C.-B. acknowledge financial support from ERC through the Starting Grant ``TESLA'' grant agreement number 805021.

\subsection*{Contributions}

M.~T. and L.~M. prepared the samples.
M.~T. performed the FIB milling and the TEM measurements.
M.~T. analyzed the TEM data.
M.~T. and S.~C.~B. prepared the figures and the discussion of the results.
L.~M. performed the structure modeling and DFT calculations.
S.C.~B. designed and supervised the experiments.
All the authors contributed to the writing of the manuscript.

\subsection*{Competing interests}

The authors declare that they have no competing interests.

\subsection*{Corresponding author}

Correspondence about this work should be sent to Sonia Conesa Boj.


\begin{thebibliography}{10}

\bibitem{ref1}
B.~Bhattacharyya, A.~Sharma, B.~Sinha, K.~Shah, S.~Jejurikar, T.~D. Senguttuvan
  et~al., {\it Evidence of robust 2d transport and Efros-Shklovskii variable
  range hopping in disordered topological insulator (Bi$_2$Se$_3$) nanowires},
  \href{https://doi.org/10.1038/s41598-017-08018-6}{{Scientific Reports}
  {\bfseries 7} (2017) 7825}.

\bibitem{ref2}
B.~Bhattacharyya, A.~Sharma, V.~P.~S. Awana, A.~K. Srivastava, T.~D.
  Senguttuvan and S.~Husale, {\it Observation of quantum oscillations in {FIB}
  fabricated nanowires of topological insulator (Bi$_2$Se$_3$)},
  \href{https://doi.org/10.1088/1361-648x/aa5536}{{Journal of Physics:
  Condensed Matter} {\bfseries 29} (2017) 115602}.

\bibitem{ref3}
K.~F. Mak, C.~Lee, J.~Hone, J.~Shan and T.~F. Heinz, {\it Atomically thin
  MoS$_2$: A new direct-gap semiconductor},
  \href{https://doi.org/10.1103/PhysRevLett.105.136805}{{Phys. Rev. Lett.}
  {\bfseries 105} (2010) 136805}.

\bibitem{ref4}
A.~Splendiani, L.~Sun, Y.~Zhang, T.~Li, J.~Kim, C.-Y. Chim et~al.,
  {\it Emerging photoluminescence in monolayer MoS$_2$},
  \href{https://doi.org/10.1021/nl903868w}{{Nano Letters} {\bfseries 10}
  (2010) 1271}.

\bibitem{ref5}
B.~Radisavljevic, A.~Radenovic, J.~Brivio, V.~Giacometti and A.~Kis,
  {\it Single-layer MoS$_2$ transistors}, {{Nature Nanotechnology}
  {\bfseries 6} (2011) 147 EP }.

\bibitem{ref6}
K.~F. Mak, K.~He, J.~Shan and T.~F. Heinz, {\it Control of valley polarization
  in monolayer MoS$_2$ by optical helicity}, {{Nature Nanotechnology}
  {\bfseries 7} (2012) 494 EP }.

\bibitem{ref7}
Y.~Li, Z.~Zhou, S.~Zhang and Z.~Chen, {\it MoS$_2$ nanoribbons: High stability
  and unusual electronic and magnetic properties},
  \href{https://doi.org/10.1021/ja805545x}{{Journal of the American
  Chemical Society} {\bfseries 130} (2008) 16739}.

\bibitem{ref8}
H.~Pan and Y.-W. Zhang, {\it Edge-dependent structural{,} electronic and
  magnetic properties of MoS$_2$ nanoribbons},
  \href{https://doi.org/10.1039/C2JM15906F}{{J. Mater. Chem.} {\bfseries
  22} (2012) 7280}.

\bibitem{ref9}
D.~Davelou, G.~Kopidakis, G.~Kioseoglou and I.~N. Remediakis, {\it MoS$_2$
  nanostructures: Semiconductors with metallic edges},
  \href{https://doi.org/https://doi.org/10.1016/j.ssc.2014.04.023}{{Solid
  State Communications} {\bfseries 192} (2014) 42 }.

\bibitem{ref10}
J.~V. Lauritsen, M.~Nyberg, R.~T. Vang, M.~V. Bollinger, B.~S. Clausen, H.~T. e
  et~al., {\it Chemistry of one-dimensional metallic edge states in
  {Mo$S_2$} nanoclusters},
  \href{https://doi.org/10.1088/0957-4484/14/3/306}{{Nanotechnology}
  {\bfseries 14} (2003) 385}.

\bibitem{ref11}
T.~P. Rossi, K.~T. Winther, K.~W. Jacobsen, R.~M. Nieminen, M.~J. Puska and
  K.~S. Thygesen, {\it Effect of edge plasmons on the optical properties of
  ${\mathrm{mos}}_{2}$ monolayer flakes},
  \href{https://doi.org/10.1103/PhysRevB.96.155407}{{Phys. Rev. B}
  {\bfseries 96} (2017) 155407}.

\bibitem{ref12}
J.~Kibsgaard, Z.~Chen, B.~N. Reinecke and T.~F. Jaramillo, {\it Engineering
  the surface structure of MoS$_2$ to preferentially expose active edge sites for
  electrocatalysis}, {{Nature Materials} {\bfseries 11} (2012) 963 EP }.

\bibitem{ref13}
D.~Voiry, M.~Salehi, R.~Silva, T.~Fujita, M.~Chen, T.~Asefa et~al.,
  {\it Conducting MoS$_2$ nanosheets as catalysts for hydrogen evolution
  reaction}, \href{https://doi.org/10.1021/nl403661s}{{Nano Letters}
  {\bfseries 13} (2013) 6222}.

\bibitem{ref14}
M.~A. Lukowski, A.~S. Daniel, F.~Meng, A.~Forticaux, L.~Li and S.~Jin,
  {\it Enhanced hydrogen evolution catalysis from chemically exfoliated
  metallic MoS$_2$ nanosheets},
  \href{https://doi.org/10.1021/ja404523s}{{Journal of the American
  Chemical Society} {\bfseries 135} (2013) 10274}.

\bibitem{ref15}
T.~F. Jaramillo, K.~P. J{\o}rgensen, J.~Bonde, J.~H. Nielsen, S.~Horch and
  I.~Chorkendorff, {\it Identification of active edge sites for
  electrochemical h2 evolution from MoS$_2$ nanocatalysts},
  \href{https://doi.org/10.1126/science.1141483}{{Science} {\bfseries 317}
  (2007) 100.}
 
\bibitem{ref16}
M.-A. Kang, S.~K. Kim, J.~K. Han, S.~J. Kim, S.-J. Chang, C.-Y. Park et~al.,
  {\it Large scale growth of vertically standing {MoS}2 flakes on 2d nanosheet
  using organic promoter},
  \href{https://doi.org/10.1088/2053-1583/aa6049}{{2D Materials}
  {\bfseries 4} (2017) 025042}.

\bibitem{ref17}
A.~V. Agrawal, N.~Kumar, S.~Venkatesan, A.~Zakhidov, C.~Manspeaker, Z.~Zhu
  et~al., {\it Controlled growth of MoS$_2$ flakes from in-plane to edge-enriched
  3d network and their surface-energy studies},
  \href{https://doi.org/10.1021/acsanm.8b00467}{{ACS Applied Nano
  Materials} {\bfseries 1} (2018) 2356}.

\bibitem{ref18}
Y.~Teng, H.~Zhao, Z.~Zhang, Z.~Li, Q.~Xia, Y.~Zhang et~al., {\it MoS$_2$
  nanosheets vertically grown on graphene sheets for lithium-ion battery
  anodes}, \href{https://doi.org/10.1021/acsnano.6b03683}{{ACS Nano}
  {\bfseries 10} (2016) 8526}.

\bibitem{ref19}
X.~Zeng, H.~Hirwa, M.~Ortel, H.~C. Nerl, V.~Nicolosi and V.~Wagner,
  {\it Growth of large sized two-dimensional MoS$_2$ flakes in aqueous solution},
  \href{https://doi.org/10.1039/C7NR00701A}{{Nanoscale} {\bfseries 9}
  (2017) 6575}.

\bibitem{ref20}
X.~Wang, H.~Feng, Y.~Wu and L.~Jiao, {\it Controlled synthesis of highly
  crystalline MoS$_2$ flakes by chemical vapor deposition},
  \href{https://doi.org/10.1021/ja4013485}{{Journal of the American
  Chemical Society} {\bfseries 135} (2013) 5304}.

\bibitem{ref21}
Y.~Jung, J.~Shen, Y.~Sun and J.~J. Cha, {\it Chemically synthesized
  heterostructures of two-dimensional molybdenum/tungsten-based dichalcogenides
  with vertically aligned layers},
  \href{https://doi.org/10.1021/nn503853a}{{ACS Nano} {\bfseries 8} (2014)
  9550}.

\bibitem{ref22}
D.~Kong, H.~Wang, J.~J. Cha, M.~Pasta, K.~J. Koski, J.~Yao et~al.,
  {\it Synthesis of MoS$_2$ and mose2 films with vertically aligned layers},
  \href{https://doi.org/10.1021/nl400258t}{{Nano Letters} {\bfseries 13}
  (2013) 1341}.

\bibitem{ref23}
H.~Wang, C.~Tsai, D.~Kong, K.~Chan, F.~Abild-Pedersen, J.~K. N{\o}rskov et~al.,
  {\it Transition-metal doped edge sites in vertically aligned MoS$_2$ catalysts
  for enhanced hydrogen evolution},
  \href{https://doi.org/10.1007/s12274-014-0677-7}{{Nano Research}
  {\bfseries 8} (2015) 566}.

\bibitem{ref24}
S.~Friedensen, J.~T. Mlack and M.~Drndi{\'c}, {\it Materials analysis and
  focused ion beam nanofabrication of topological insulator bi2se3},
  \href{https://doi.org/10.1038/s41598-017-13863-6}{{Scientific Reports}
  {\bfseries 7} (2017) 13466}.

\bibitem{ref25}
P.~A. Sharma, A.~L. Lima~Sharma, M.~Hekmaty, K.~Hattar, V.~Stavila, R.~Goeke
  et~al., {\it Ion beam modification of topological insulator bismuth
  selenide}, \href{https://doi.org/10.1063/1.4904936}{{Applied Physics
  Letters} {\bfseries 105} (2014) 242106.}

\bibitem{ref26}
D.~S. Fox, Y.~Zhou, P.~Maguire, A.~O'Neill, C.~{\'O}'Coile{\'a}in, R.~Gatensby
  et~al., {\it Nanopatterning and electrical tuning of MoS$_2$ layers with a
  subnanometer helium ion beam},
  \href{https://doi.org/10.1021/acs.nanolett.5b01673}{{Nano Letters}
  {\bfseries 15} (2015) 5307}.

\bibitem{ref27}
B.~Yue, F.~Hong, K.-D. Tsuei, N.~Hiraoka, Y.-H. Wu, V.~M. Silkin et~al.,
  {\it High-energy electronic excitations in a bulk
  $\mathrm{Mo}{\mathrm{s}}_{2}$ single crystal},
  \href{https://doi.org/10.1103/PhysRevB.96.125118}{{Phys. Rev. B}
  {\bfseries 96} (2017) 125118}.

\bibitem{ref28}
L.~Martin, R.~Mamy, A.~Couget and C.~Raisin, {\it Optical properties and
  collective excitations in MoS$_2$ and nbse2 in the 1.7 to 30 ev range},
  \href{https://doi.org/10.1002/pssb.2220580223}{{physica status solidi
  (b)} {\bfseries 58} (1973) 623.}
  

\bibitem{ref29}
D.~N. Basov, M.~M. Fogler and F.~J. Garc{\'\i}a~de Abajo, {\it Polaritons in
  van der Waals materials},
  \href{https://doi.org/10.1126/science.aag1992}{{Science} {\bfseries 354}
  (2016).}

\bibitem{ref30}
F.~Tran, J.~Stelzl, D.~Koller, T.~Ruh and P.~Blaha, {\it Simple way to apply
  nonlocal van der Waals functionals within all-electron methods},
  \href{https://doi.org/10.1103/PhysRevB.96.054103}{{Phys. Rev. B}
  {\bfseries 96} (2017) 054103}.

\bibitem{ref31}
P.~Blaha, K.~Schwarz, G.~K.~H. Madsen, D.~F. Kvasnicka and J.~Luitz,
  {\it Wien2k: An augmented plane wave plus local orbitals program for
  calculating crystal properties},  2001.

\bibitem{ref32}
J.~Klime{\v{s}}, D.~R. Bowler and A.~Michaelides, {\it Chemical accuracy for
  the van der Waals density functional},
  \href{https://doi.org/10.1088/0953-8984/22/2/022201}{{Journal of
  Physics: Condensed Matter} {\bfseries 22} (2009) 022201}.

\bibitem{ref33}
J.~c.~v. Klime\ifmmode~\check{s}\else \v{s}\fi{}, D.~R. Bowler and
  A.~Michaelides, {\it Van der Waals density functionals applied to solids},
  \href{https://doi.org/10.1103/PhysRevB.83.195131}{{Phys. Rev. B}
  {\bfseries 83} (2011) 195131}.

\bibitem{ref34}
A.~D. Becke, {\it Density-functional exchange-energy approximation with
  correct asymptotic behavior},
  \href{https://doi.org/10.1103/PhysRevA.38.3098}{{Phys. Rev. A}
  {\bfseries 38} (1988) 3098}.

\bibitem{ref35}
J.~P. Perdew and Y.~Wang, {\it Accurate and simple analytic representation of
  the electron-gas correlation energy},
  \href{https://doi.org/10.1103/PhysRevB.45.13244}{{Phys. Rev. B}
  {\bfseries 45} (1992) 13244}.

\bibitem{ref36}
M.~Dion, H.~Rydberg, E.~Schr\"oder, D.~C. Langreth and B.~I. Lundqvist,
  {\it Van der Waals density functional for general geometries},
  \href{https://doi.org/10.1103/PhysRevLett.92.246401}{{Phys. Rev. Lett.}
  {\bfseries 92} (2004) 246401}.

\bibitem{ref37}
P.~Blöchl, O.~Jepsen and O.~Andersen, {\it Improved tetrahedron method for
  brillouin-zone integrations},
  \href{https://doi.org/10.1103/physrevb.49.16223}{{Physical review. B,
  Condensed matter} {\bfseries 49} (1994) 16223—16233}.

\bibitem{ref38}
B.~Schönfeld, J.~J. Huang and S.~C. Moss, {\it Anisotropic mean-square
  displacements (msd) in single-crystals of 2h- and 3r-MoS$_2$},
  \href{https://doi.org/10.1107/S0108768183002645}{{Acta Crystallographica
  Section B} {\bfseries 39} (1983) 404}.

\end{thebibliography}

\providecommand{\href}[2]{#2}\begingroup\raggedright\endgroup

\end{document}